\documentclass{article}

\usepackage{algorithm}
\usepackage{algpseudocode}
\usepackage{amsthm}

\usepackage{comment}
\usepackage{mathtools}
\usepackage{subcaption}
\usepackage{multirow}
\usepackage{comment}
\usepackage[authoryear]{natbib}

\let\temp\rmdefault
\usepackage{mathpazo}
\let\rmdefault\temp

\allowdisplaybreaks
\usepackage{color}
\usepackage{colortbl}
\usepackage[colorlinks, allcolors=blue]{hyperref}
\usepackage{hyperref}
\usepackage[british]{babel}
\usepackage{diagbox}
\usepackage{rotating}
\usepackage[justification=centering]{caption}
\theoremstyle{definition}
\newtheorem{Def}{Definition}[section]
\newtheorem{Bsp}{Example}[section]

\newtheorem{remark}[Def]{Remark}
\theoremstyle{plain}

\newtheorem{Thm}[Def]{Theorem}
\newtheorem{Prop}[Def]{Proposition}

\newtheorem{As}[Def]{Assumption}
\theoremstyle{remark}

\numberwithin{equation}{section}

\newcommand{\norm}[1]{\left\lVert#1\right\rVert}

\newcommand{\N}{\mathbb{N}}
\newcommand{\R}{\mathbb{R}}

\newcommand{\E}{\mathbb{E}}
\newcommand{\PR}{\mathbb{P}}

\newcommand{\ve}{\varepsilon}

\newcommand{\hl}{\hat{\lambda}}

\newcommand{\hs}{\hat{\Sigma}}
\newcommand{\hks}{\hat{\sigma}}

\renewcommand{\algorithmicrequire}{\textbf{Input:}}
\renewcommand{\algorithmicensure}{\textbf{Output:}}

\usepackage[T1]{fontenc}
\usepackage[
left=2.00cm,
right=2.00cm,
top=2.00cm,
bottom=2.00cm
]{geometry}
\usepackage{lipsum}

\begin{document}

\twocolumn[

\title{Multivariate Mean Comparison under Differential Privacy}

\author{ Martin Dunsche, Tim Kutta,  Holger Dette }
\date{%
    Ruhr-University Bochum
}
\maketitle
]

\begin{abstract}

\noindent The comparison of multivariate population means is a central task of statistical inference. While statistical theory provides a variety of analysis tools, they usually do not protect individuals' privacy. This knowledge can create incentives for participants in a study to conceal their true data (especially for outliers),
which might result in a distorted analysis. 
In this paper we address this problem by developing
a hypothesis test for multivariate mean comparisons that guarantees differential privacy to users. The test statistic is based on the popular Hotelling's $t^2$-statistic, which has a natural interpretation in terms of the Mahalanobis distance. In order to control the type-1-error, we present a bootstrap algorithm under differential privacy that provably yields a reliable test decision. In an empirical study we demonstrate the applicability of this approach.

\end{abstract}

\section{Introduction}

Over the last decades the availability of large data bases has transformed statistical practice. While data mining flourishes, users are concerned about increasing transparency vis-a-vis third parties. To address this problem new analysis tools have been devised that balance precise inference with solid privacy guarantees. \\
In this context, statistical tests that operate under \textit{differential privacy} (DP) are of interest: Statistical tests are the standard tool to validate hypotheses regarding data samples and to this day form the spine of most empirical sciences. Performing tests under DP means determining general trends in the data, while masking individual contribution. This makes it hard for adversaries to retrieve unpublished, personal information from the published   analysis.
\\\textbf{Related works:} In recent years hypothesis testing under DP has gained increasing attention.
In a seminal work \cite{smith2011privacy} introduces a privatization method,  for a broad class of test statistics, that guarantees DP without impairing asymptotic performance. Other theoretical aspects such as optimal tests under DP are considered in  \cite{canonne2019structure}. Besides such theoretical investigations, a number of privatized tests have been devised to replace classical inference, where sensitive data is at stake.
For example \cite{gaboardi2016differentially} and \cite{pmlr-v54-rogers17a} consider privatizations of classical goodness of fit tests for categorical data, tailored to applications in genetic research, where privacy of study participants is paramount. In a closely related work,  \cite{Wang2015} use privatized likelihood-ratio statistics to validate various assumptions for tabular data. Besides, \cite{sei2021privacy} propose a method for privatizations in small sample regimes. \\
A cornerstone of statistical analysis is the study of population means and accordingly this subject has attracted particular attention.
For example, \cite{ding2018comparing} develop a  private t-test to compare population means under local differential privacy, \cite{karwa2017finite} and \cite{du2020differentially} construct private confidence intervals for the mean (which is equivalent to the one-sample t-test) under global DP and \cite{swanberg2019improved} suggests a differentially private ANOVA. Moreover, \cite{couch2019differentially} present privatizations for a number of non-parametric tests (such as Wilcoxon signed-rank tests) and \cite{ferrando2020general} devise general confidence intervals for exponential families.  \\
A key problem of statistical inference under DP consists in the fact that privatization  inflates the variance of the test statistics. If this is not taken into account  properly, it can destabilize subsequent analysis and lead to the "discovery" of spurious effects.  To address these problems, recent works (such as \cite{gaboardi2016differentially} and \cite{ferrando2020general}) have employed resampling procedures that explicitly incorporate the effects of privatization and are therefore more reliable than tests 
based on standard, asymptotic theory.
\\\textbf{Our contributions:} In this work we present a test for multivariate mean comparisons under DP, based on the popular Hotelling's $t^2$-statistic. We retrieve the effect that asymptotic test decisions work under DP, as long as privatizations are weak, whereas for strong privatizations, they yield distorted results (see Section \ref{Sec_4} for details). As a remedy we consider a parametric bootstrap that cuts false rejections and is provably consistent for increasing sample size. This method can be extended to other testing problems,
is easy to implement (even for non-expert users) and can be efficiently automatized as part of larger data disseminating structures. We demonstrate the efficacy of our approach, even for higher dimensions and strong privatizations, in a simulation study. 
The proofs of all mathematical results are deferred to the Appendix.

\section{Mathematical background}

In this Section we provide the mathematical context for private mean comparisons, beginning with a general introduction into two sample tests. Subsequently we discuss  Hotelling's $t^2$-test, which is a standard tool to assess mean deviations. Finally, we define the notion of differential privacy and consider key properties, such as stability under post-processing. Readers familiar with any of these topics can skip the respective section.

\subsection{Statistical tests for two samples} \label{Sec_31}

In this work we are interested in testing statistical hypotheses regarding the distribution of two data samples (of random vectors) $X_1,...,X_{n_1}$ and $Y_1,...,Y_{n_2}$. 
\\
Statistical tests are decision rules that select one out of two rivaling hypotheses $H_0$ and $H_1$, where $H_0$ is referred to as the "null hypothesis" (default belief) and $H_1$ as the "alternative". To make this decision, a statistical test creates a summary statistic $S:=S(X_1,...,X_{n_1}, Y_1,...,Y_{n_2})$ from the data and based on $S$ determines whether to keep $H_0$, or to switch to $H_1$. Typically, the decision to reject $H_0$ in favor of $H_1$ is made, if $S$ surpasses a certain threshold $q$, above which, the value of $S$ seems at odds with $H_0$. 
In this situation the threshold $q$ may or may not depend on the data samples. \\
Given the randomness in statistical data, there is always a risk of making the wrong decision. Hypothesis-alternative-pairs $(H_0, H_1)$ are usually formulated such that mistakenly keeping $H_0$ inflicts only minor costs on the user, while wrongly switching to $H_1$ produces major ones. In this spirit, tests are constructed to keep the risk of false rejection below a predetermined level $\alpha$, i.e. $
\PR_{H_0}(S>q)\leq \alpha$, which is referred to as the \textit{nominal level}  (or \textit{type-1-error}). Commonly the nominal level is chosen as $\alpha \in \{0.1, 0.05, 0.01\}$. Notice that $\alpha$ can be regarded as an input parameter of the threshold $q=q(\alpha)$. Even though sometimes an exact nominal level can be guaranteed, in practice most tests only satisfy an asymptotic nominal level, i.e.
\begin{equation*}
    \limsup_{n_1,n_2 \to \infty }\mathbb{P}_{H_0}(S >q(\alpha)) \leq \alpha.
\end{equation*}
Besides controlling the type-1-error, a reasonable test has to be \textit{consistent}, i.e. it has to reject $H_0$ if $H_1$ holds and sufficient data is available. In terms of the summary statistic $S$, this means that $S$ increases for larger data samples and transgresses $q(\alpha)$ with growing probability
\begin{equation*}
    \lim_{n_1,n_2 \to \infty }\mathbb{P}_{H_1}(S >q(\alpha) ) = 1.
\end{equation*}

\subsection{Hotelling's $t^2$-test}

We now consider a specific test for the comparison of multivariate means: Suppose that two independent samples of random vectors $X_1,...,X_{n_1}$ and $Y_1,...,Y_{n_2}$ are given, both stemming from the $d$-dimensional cube $[-m,m]^d$, where $m>0$ and $d \in \mathbb{N}$. Furthermore we assume that both samples consist of independent identically distributed (i.i.d) observations. Conceptually each vector corresponds to the data of one individual and we want to use these to test the "hypothesis-alternative"-pair
	\begin{equation} \label{Eq_hypothesis}
		H_0: \mu_X=\mu_Y~, \quad \quad H_1: \mu_X \neq \mu_Y~,
	\end{equation}
where $\mu_X:=\E[X_1]\in \R^d,\mu_Y:=\E[Y_1] \in \R^d$ denote the respective expectations.
A standard way to test \eqref{Eq_hypothesis} is provided by \textit{Hotelling's $t^2$-test}, which is based on the test statistic
\begin{equation}\label{Hot}
		t^2= \frac{n_1n_2}{n_1+n_2} (\bar{X}-\bar{Y})^T\hat{\Sigma}^{-1}(\bar{X}-\bar{Y})~,
\end{equation}
where $\bar X= \frac{1}{n_1}\sum_{i=1}^{n_1}X_i$ and $\bar Y=\frac{1}{n_2}\sum_{i=1}^{n_2}Y_i$ denote the respective sample means and the pooled sample covariance is given by
\begin{equation*}
		\hat{\Sigma}=\frac{(n_1-1)\hat{\Sigma}_{X}+(n_2-1)\hat{\Sigma}_{Y}}{n_1+n_2-2}. 
\end{equation*}
Here $\hat{\Sigma}_X=\frac{1}{n_1-1}\sum_{i=1}^{n_1} (X_i-\mu_{X})(X_i-\mu_{X})^\top$ and $\hat{\Sigma}_Y=\frac{1}{n_2-1}\sum_{i=1}^{n_2} (Y_i-\mu_{Y})(Y_i-\mu_{Y})^\top$ denote the sample covariance matrices of $X_1$ and $Y_1$, respectively. Assuming that $\Sigma_X=\Sigma_Y$ (a standard condition for Hotelling's $t^2$-test) $\hat{\Sigma}$ is a consistent estimator for the common covariance.\\
We briefly formulate a few observations regarding the $t^2$-statistic:
\begin{itemize}
    \item[i)] In the simple case of $d=1$, the $t^2$-statistic collapses to the (squared) statistic of the better-known two sample t-test.
    \item[ii)] We can rewrite 
    $$
    t^2 = \frac{n_1n_2}{n_1+n_2} \norm{\hat{\Sigma}^{-1/2}(\bar{X}-\bar{Y})}_2^2.
    $$
    As a consequence, the $t^2$-statistic is non-negative and assumes high values if $\bar{X}-\bar{Y} \approx \mu_X - \mu_Y$ is large in the norm.
    \item[iii)] The $t^2$-statistic is closely related to the Mahalanobis distance, which is a standard measure for multivariate mean comparisons (see \cite{DEMAESSCHALCK20001}).
\end{itemize}

In order to formulate a statistical test based on the $t^2$-statistic, we consider its large sample behavior. Under the hypothesis $\sqrt{n_1n_2/(n_1+n_2)}\hat{\Sigma}^{-1/2}(\bar{X}-\bar{Y})$ follows (approximately) a $d$-dimensional, standard normal distribution, such that its squared norm (that is the $t^2$-statistic) is approximately $\chi_d^2$ distributed (chi-squared with $d$ degrees of freedom).   
Now if  $q_{1-\alpha}$ denotes the upper $\alpha$-quantile of the $\chi_d^2$ distribution, the test decision "reject $H_0$ if $t^2> q_{1-\alpha}$", yields a consistent, asymptotic level $\alpha$-test for any $\alpha \in (0,1)$. For 
details on Hotelling's $t^2$-test we refer to \cite{mardia}.

\subsection{Differential privacy} \label{Sec_33}

Differential privacy (DP) has over the last decade become the de facto gold standard in privacy assessment of data disseminating procedures (see e.g. \cite{census}, \cite{10.1145/2660267.2660348} or \cite{rogers2020linkedin}). Intuitively DP describes the difficulty of inferring individual inputs from the releases of a randomized algorithm. This notion is well suited to a statistical framework, where a trusted institution, like a hospital, publishes results of a study (algorithmic releases), but candidates would prefer to conceal participation (individual inputs). To make this notion mathematically rigorous, we consider data bases $\mathbf{x}, \mathbf{x}' \in \mathcal{D}^n$, where $\mathcal{D}$ is some set, and call them \textit{adjacent} or \textit{neighboring}, if they differ in only one entry. 

\begin{Def}
A randomized algorithm $A: \mathcal{D}^n \to \mathbb{R}$ is called $\ve$\textit{-differentially private} for some $\ve>0$, if for any measurable event $E \subset \R$ and any adjacent $\mathbf{x}, \mathbf{x}'$
\begin{equation} \label{Eq_def_DP}
    \mathbb{P}(A(\mathbf{x})\in E) \le e^\ve \,\, \mathbb{P}(A(\mathbf{x}')\in E)
\end{equation}
holds. 
\end{Def}

Condition \eqref{Eq_def_DP} requires that the distribution of $A(\mathbf{x})$ does not change too much, if one entry of $\mathbf{x}$ is exchanged (where small $\ve$ correspond to less change and thus stronger privacy guarantees). In statistical applications, private algorithms are usually assembled in a modular fashion: They take as building blocks some well-known private algorithms (e.g. the Laplace or Exponential Mechanism), use them to privatize key variables (empirical mean, variance etc.) and aggregate the privatized statistic. This approach is justified by two stability properties of DP: Firstly, privacy preservation under post-processing, which ensures that if $A$ satisfies $\ve$-DP, so does any measurable transformation $h(A)$. Secondly, the composition theorem that maintains at least $\sum_{i=1}^k \ve_i$-DP of a vector $(A_1,...,A_k)$ of algorithms, where $A_i$ are independent $\ve_i$-differentially private algorithms. In the next Section we employ such a modular privatization of the Hotelling's $t^2$-statistic for private mean comparison. We conclude our discussion on privacy with a small remark on the role of the "trusted curator".
\begin{remark}
Discussions of (global) DP usually rely on the existence of some "trusted curator" who aggregates and privatizes data before publication. In reality this role could be filled by an automatized, cryptographic protocol (secure multi-party computation), which calculates and privatizes the statistic before publication without any party having access to the full data set (for details see \cite{lindell2005secure}, \cite{bogetoft2009secure}). This process has the positive side-effect that it prevents a curator from re-privatizing if an output seems too outlandish (overturning privacy in the process).
\end{remark}

\section{Privatized mean comparison} \label{Sec_3}

In this Section we introduce a privatized version  $t^{DP}$ of Hotelling's $t^2$-statistic. Analogous to the traditional $t^2$-statistic, the rejection rule "$t^{DP}>q_{1-\alpha}$" yields in principle a consistent, asymptotic level-$\alpha$ test for $H_0$ (see Theorem \ref{Theorem_asymptotic}). However, empirical rejection rates often exceed the prescribed nominal level $\alpha$ for a combination of low sample sizes and high privatization (see Example \ref{Example_2}).

As a consequence we devise a parametric bootstrap for a data-driven rejection rule. We validate this approach theoretically (Theorem \ref{Theorem_bootstrap}) and demonstrate empirically a good approximation of the nominal level in Section \ref{Sec_4}.

\subsection{Privatization of the $t^2$-statistic}

We begin this Section by formulating the Assumptions of the following, theoretical results:
\begin{As}$ $\\[-5ex] \label{Assumption_1}
\begin{itemize}
    \item[(1)] The samples $X_1,\hdots, X_{n_1}$ and $Y_1,\hdots, Y_{n_2}$ are independent, each consisting  of i.i.d. observations and are both supported on the cube $[-m,m]^d$, for some known $m>0$.
    \item[(2)] The covariance matrices 
    \begin{align*}
        \Sigma_X:= &\mathbb{E}[(X_1-\mu_{X})(X_1-\mu_{X})^T] \\
        \Sigma_Y: =& \mathbb{E}[(Y_1-\mu_{Y})(Y_1-\mu_{Y})^T]
    \end{align*}
    are identical and invertible.
    \item[(3)] The sample sizes $n_1,n_2$ are of the same order. That is with $n:=n_1+n_2$ we have
    \begin{equation*}
        \lim_{n \to \infty} \frac{n_i}{n} =\xi_i \in (0,1) \quad \quad i=1,2.
    \end{equation*}

\end{itemize}
\end{As}

We briefly comment on the Assumptions made.
\begin{remark}
(1): The assumption of independent observations is common in the literature on machine learning and justified in many instances. Boundedness of the data -with some known bound- is an important precondition for standard methods of privatization 
(such as the below discussed Laplace Mechanism or the ED algorithm). Generalization are usually possible (see e.g. \cite{smith2011privacy}) but lie beyond the scope of this paper. \\
(2): Invertibility of the covariance matrices is necessary to define the Mahalanobis distance. If this assumption is violated either using another distance measure (defining a different test) or a prior reduction of dimensions is advisable.
\\
Equality of the matrices $\Sigma_X = \Sigma_Y$ is assumed for ease of presentation, but can be dropped, if the pooled estimate $\hat \Sigma$ is replaced by the re-weighted version 
$$
    \hat \Sigma^{\neq} := \frac{n_2 \hat \Sigma_X + n_1 \hat \Sigma_Y }{n_1+n_2}.
$$
(3): We assume that asymptotically the size of each group is non-negligible. This assumption is standard in the analysis of two sample tests and implies that the noise in the estimates of both groups is of equal magnitude. If this was not the case and e.g. $\xi_1=0$ (in practice $n_1<<n_2$) it is more appropriate to model the situation as a one-sample test (as $\mu_{Y}$ is basically known).
\end{remark}
Recall the definition of  Hotelling's $t^2$-statistic in \eqref{Hot}. By construction we can express the $t^2$-statistic as a deterministic function of four data dependent entities: The sample means $\bar{X}, \bar{Y}$ and the sample covariance matrices $\hat{\Sigma}_X, \hat{\Sigma}_Y$. According to the \textit{composition-} and \textit{post-processing theorem} of DP (see Section \ref{Sec_33}) we can privatize the $t^2$-statistic by privatizing each of these inputs. \\
For the privatization of the sample means, we use the popular \textit{Laplace Mechanism} (see \cite{10.1561/0400000042}, p.32): It is well-known  that $\bar X^{DP} :=\bar{X}+Z$ and $\bar Y^{DP} :=\bar{Y}+Z'$ fulfill $\ve/4$-DP, if $Z=(Z_1,\hdots,Z_d)^T$ and $Z'=(Z_1',\hdots,Z_d')^T$ consist of independent random variables $Z_k \sim Lap(0, \frac{2 md}{n_1 (\ve/4)})$ and $Z_k' \sim Lap(0, \frac{2 md}{n_2 (\ve/4)})$ for $k=1,...,d$. \\
For the privatization of the covariance matrices $\hat{\Sigma}_X, \hat{\Sigma}_Y$ we employ the $ED$ Mechanism, specified in the Appendix (which is a simple adaption of the Algorithm proposed in \cite{NEURIPS2019_4158f6d1}). We can thus  define
differentially private estimates  $\hat{\Sigma}_X^{DP}:=ED(\hat \Sigma_X, \ve/4)$ and $\hat{\Sigma}_Y^{DP}:=ED(\hat \Sigma_Y, \ve/4)$, both satisfying $\ve/4$-DP. We point out that the outputs of $ED$ are always covariance matrices (positive semi-definite and symmetric).
Therewith, we can define a privatized pooled sample covariance matrix as
\begin{align}\label{pooled_priv}
	\hat \Sigma^{DP}:= &\frac{(n_1-1)\hat{\Sigma}_X^{DP}+(n_2-1)\hat{\Sigma}_Y^{DP}}{n_1+n_2-2} \\
	& + diag(c_1+c_2)~, \nonumber 
\end{align} 
where $c_1:=2(\frac{2md}{n_1(\ve/4)})^2, c_2:=2(\frac{2md}{n_2(\ve/4)})^2$ are corrections accounting for variance increase, due to the mean privatizations.
\begin{algorithm}[h]
	 \small
    \algorithmicrequire \; \parbox[t]{\dimexpr\linewidth-\algorithmicindent}{means: $\bar{X}$, $\bar{Y}$, covariance matrices: $\hat{\Sigma}_X$, $\hat{\Sigma}_Y$, \\
    privacy level: $\ve$} \\[0.2cm]
    \algorithmicensure \, $\bar{X}^{DP}$, $\bar{Y}^{DP}$, $\hat{\Sigma}_X^{DP}$, $\hat{\Sigma}_Y^{DP}$
    \begin{algorithmic}[1]
    \Function{PS}{$\bar{X}$, $\bar{Y}$, $\hat{\Sigma}_X$, $\hat{\Sigma}_Y$, $\varepsilon$}
	\For {$i=1,\hdots,d$}
    \State Generate $Z_i \sim Lap(0, \frac{2 md}{n_1 \ve/4})$
    \State Generate $Z_i' \sim Lap(0, \frac{2 md}{n_2 \ve/4})$
    \EndFor
    \State Set $\bar{X}^{DP} := \bar{X}+(Z_1,...,Z_d)$, $\bar{Y}^{DP} := \bar{Y}+(Z_1',...,Z_d')$
    \State Set $\hat{\Sigma}_X^{DP} = ED(\hat \Sigma_X, \varepsilon/4)$, $\hat{\Sigma}_Y^{DP} = ED(\hat\Sigma_Y, \varepsilon/4)$
    \State \Return $\bar{X}^{DP},\bar{Y}^{DP},\hat{\Sigma}_X^{DP},\hat{\Sigma}_Y^{DP}$
    \EndFunction
    \end{algorithmic}
	\caption{Privatized statistics (PS)}\label{alg_privatization}
\end{algorithm}
Finally we can formulate a privatized version of the Hotelling's $t^2$-statistic as follows:
\begin{align}\label{test_priv}
		t^{DP}= & \frac{n_1n_2}{n_1+n_2} (\bar{X}^{DP}-\bar{Y}^{DP})^T[\hat{\Sigma}^{DP}]^{-1}\\
		& \quad\quad  \,\, \times (\bar{X}^{DP}-\bar{Y}^{DP})~ \nonumber \\
		= &  \frac{n_1n_2}{n_1+n_2} \norm{[\hat{\Sigma}^{DP}]^{-1/2}(\bar{X}^{DP}-\bar{Y}^{DP})}_2^2 \nonumber
\end{align}

\begin{Thm}\label{privacy1}
	The privatized $t^2$-statistic $t^{DP}$ is $\ve$-\textit{differentially private}.
\end{Thm}

In the one dimensional case, the covariance privatization by $ED$ boils down to an application of the Laplace Mechanism and $t^{DP}$ has a simple closed form.
\begin{Bsp}[Privatization in $d=1$]\label{Example d=1}
Assume that $d=1$. Then the data $X_1,\hdots,X_{n_1}$ and $Y_1,\hdots, Y_{n_2}$ originates from the interval $[-m,m]$ and we can write the privatized test statistic as
\begin{align*}
    	t^{DP}=& 
    	\frac{n_1n_2}{n_1+n_2} \frac{(\bar{X}^{DP}-\bar{Y}^{DP})^2}{(\sigma^{DP})^2}~,
\end{align*}
where 
\begin{align*}
	(\sigma^{DP})^2:=& \frac{(n_1-1)(|\hks_X+L_1|)+(n_2-1)(|\hks_Y+L_2|)}{n_1+n_2-2}\\& + 2 \Big(\frac{2m}{n_1(\ve/4)}\Big)^2+2\Big(\frac{2m}{n_2(\ve/4)}\Big)^2~.
	\end{align*}
Here, $L_1$ and $L_2$ follow a centered Laplace distribution, with variance specified in the Appendix. Note that the privatization of $\hks_X^2$ and $\hks_Y^2$ is conform with the privatization of Algorithm $ED$ (see Appendix), since the first (and only) eigenvalue is the sample variance itself, while privatization of eigenvectors is a non-issue for $d=1$.
\end{Bsp}

As for the non-privatized $t^2$-statistic, we can prove under $H_0$ that $t^{DP}$ approximates a $ \chi^2_d$-distribution as $n_1, n_2 \to \infty$.
This means that (at least for large sample sizes) the perturbations introduced by the Laplace noise and the $ED$-algorithm are negligible. 
\begin{algorithm}
	 \small
    \algorithmicrequire \; \parbox[t]{\dimexpr\linewidth-\algorithmicindent}{means: $\bar{X}^{DP}$, $\bar{Y}^{DP}$, covariance matrices: $\hat{\Sigma}_X^{DP}$,\\ $\hat{\Sigma}_Y^{DP}$, quantile: $q$} \\[0.2cm]
    \algorithmicensure \, $choice \in \{0,1\}$ coding for acceptation ($0$) or rejection ($1$) of $H_0$ 
    \begin{algorithmic}[1]
    \Function{PHT}{$\bar{X}^{DP}$, $\bar{Y}^{DP}$, $\hat{\Sigma}_X^{DP}$, $\hat{\Sigma}_Y^{DP}$, $q$}
    \State Compute $t^{DP}$ (defined in \ref{test_priv})
    \State Define $choice=0$
    \If{$t^{DP}>q$}
        \State Set $choice=1$
    \EndIf
    \State \Return $choice$
    \EndFunction
    \end{algorithmic}
	\caption{Privatized Hotelling's $t^2$-test (PHT)}\label{alg_test}
\end{algorithm}
\begin{Thm}\label{Theorem_asymptotic}
    The decision rule "reject if 
    \begin{equation}  \label{hd1}
t^{DP}>q_{1-\alpha}
    \end{equation}
   (Algorithm \ref{alg_test})" where  $q=q_{1-\alpha}$ is   the ($1-\alpha$)-quantile of $\chi_d^2$ distribution,
   yields a consistent, asymptotic level-$\alpha$ test for the hypotheses \eqref{Eq_hypothesis}.
\end{Thm}

Theorem \ref{Theorem_asymptotic} underpins the assertion that "asymptotically privatizations do not matter". Yet in practice, privatizations can have a dramatic impact on the (finite sample) performance of tests.

\begin{figure}[H]
\begin{subfigure}{\columnwidth}
\centering
\includegraphics[width=\linewidth]{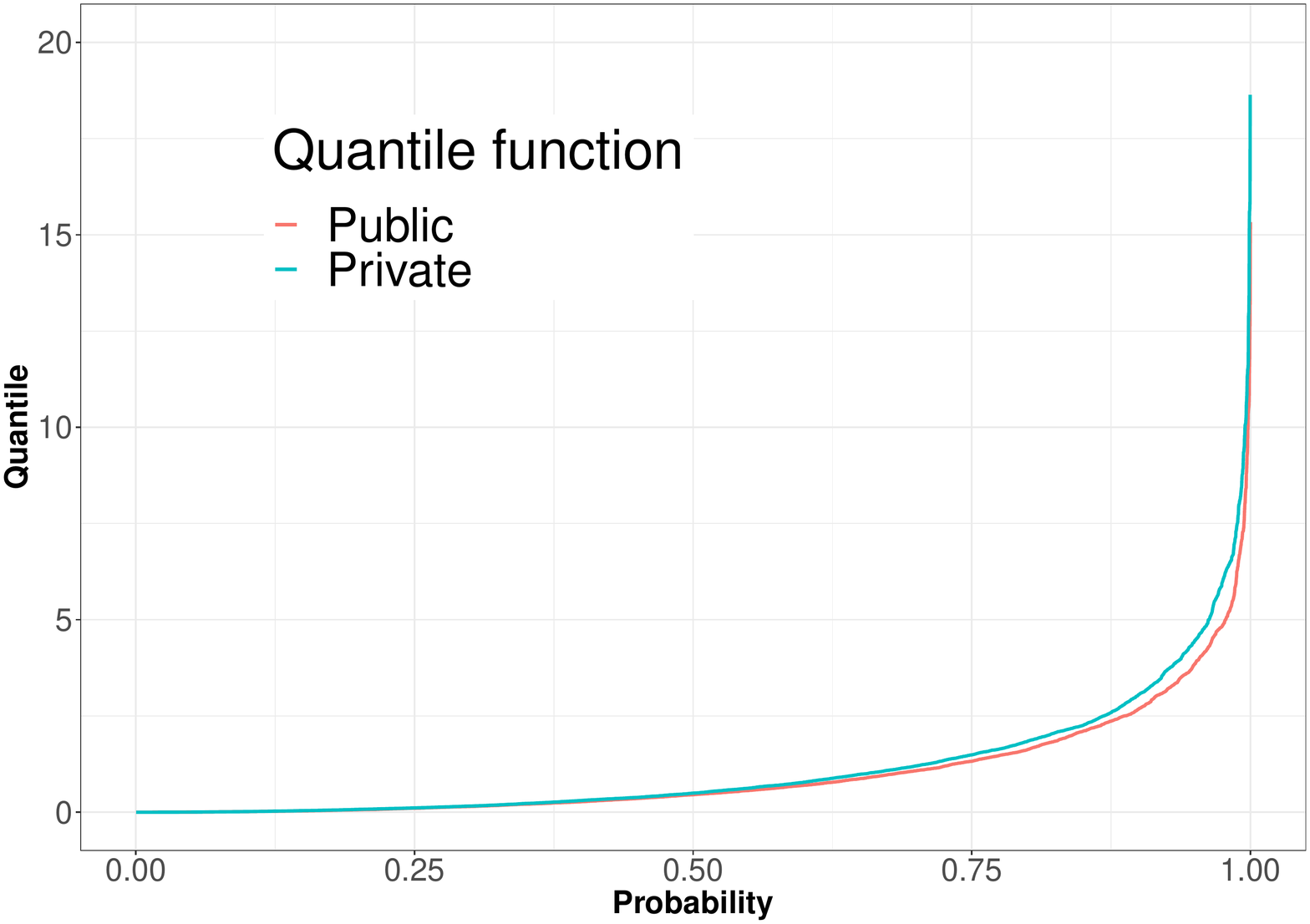}
\end{subfigure}
 
\begin{subfigure}{1\columnwidth}
\centering
\includegraphics[width=\linewidth]{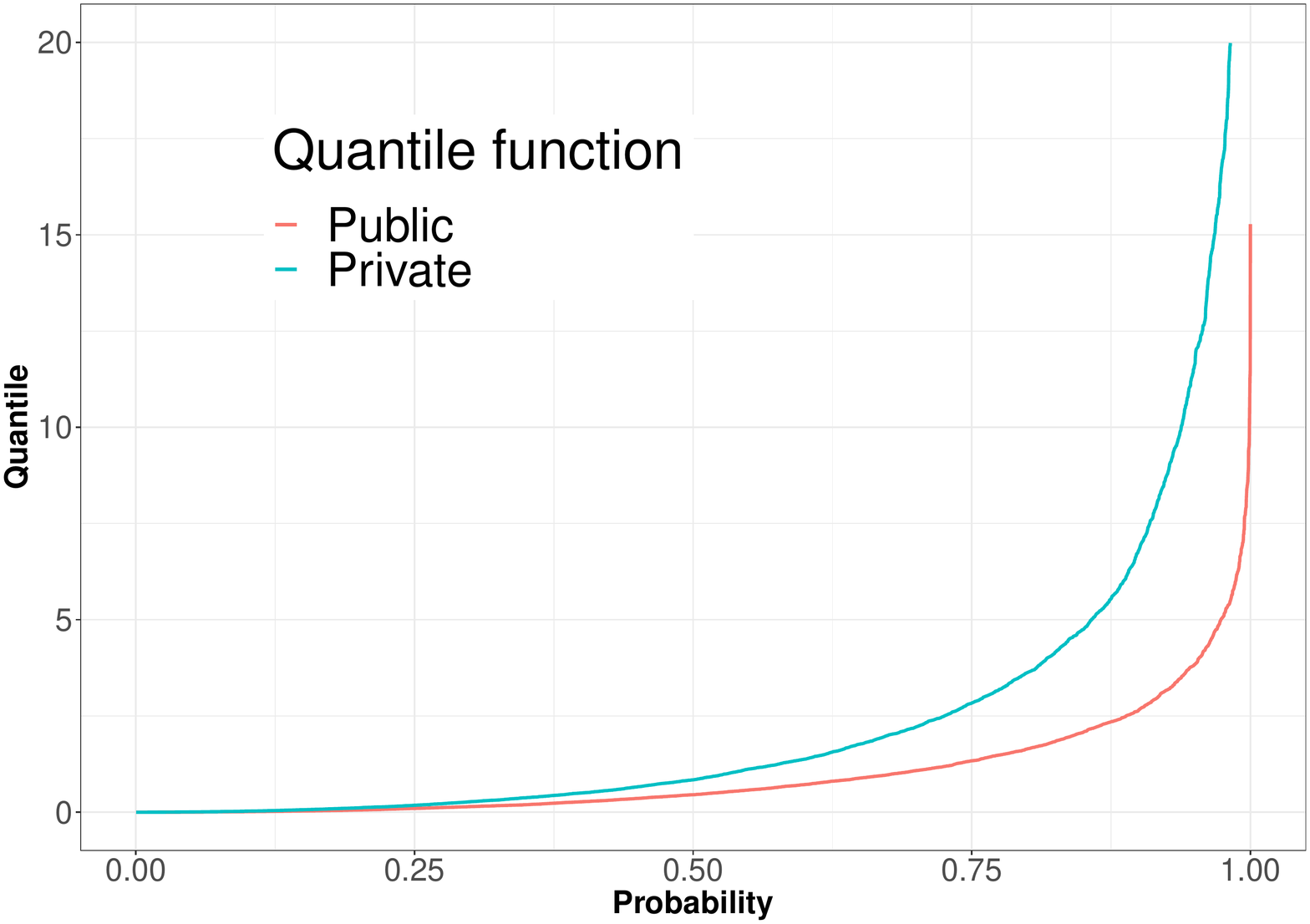}
\end{subfigure}
 \caption{Simulated quantile functions for $\chi^2_1$ (red) and $t^{DP}$ (blue) for privacy levels $\varepsilon=4$ (top) and $\varepsilon=1$ (bottom).\label{Fig_quantile_functions}}
\end{figure}

\begin{Bsp}[Effects of privatization]  \label{Example_2}
In most instances, privatizing a test statistic has no influence on its asymptotic behavior, s.t. rejection rules based on asymptotic quantiles remain theoretically valid. However, empirical studies demonstrate, that in practice even moderate privacy levels can lead to inflated type-$1$-errors -- in our case because the quantiles of the $\chi^2_d$-distribution do not provide good approximations for those of $t^{DP}$.\\
To illustrate this effect we consider the case $d=1$, discussed in Example \ref{Example d=1} for samples of sizes $n_1=n_2=500$, both of which drawn according to the same density, $f(t) \propto \exp(-2 t^2)$ on the interval $[-1,1]$. We simulate the quantile
functions (inverse of the distribution function) of $\chi_1^2$ and $t^{DP}$ respectively for privacy levels $\varepsilon=1,4$. Figure \ref{Fig_quantile_functions} indicates that for moderate privacy guarantees ($\varepsilon=4$) the distribution of $t^{DP}$ is close to that of the $\chi_1^2$, s.t. for instance 
$\mathbb{P}_{H_0}(t^{DP} >q_{0.95}) \approx 6.8 \%$ (where again $q_{1-\alpha}$ is the $\alpha$ quantile of the $\chi_1^2$-distribution).
 This approximation seems reasonable, but it deteriorates quickly for smaller $\varepsilon$. Indeed for $\varepsilon=1$  we observe that that $\mathbb{P}_{H_0}(t^{DP} >q_{0.95}) \approx 18.9 \%$, which     is a dramatic error. This effect is still more pronounced in higher dimensions and much larger sample sizes are needed to mitigate it (for details see Table \ref{tab:1}).
\end{Bsp}

Summarizing this discussion, we recommend to use 
Hotelling's $t^2$-test \eqref{hd1}  based on the privatized statistic $t^{DP}$
with the standard (asymptotic) quantiles only 
in situations where sample sizes are large, the dimension is small and  privatizations are weak. In all other cases, specifically for larger dimension and stronger privatization, the quantiles have to be adapted to avoid inflated rejection errors under the null hypothesis.

\subsection{Bootstrap}

In this Section we consider a modified rejection rule for $H_0$, based on $t^{DP}$, that circumvents the problem of inflated type-$1$-error  (see Example \ref{Example_2}). Privatizations increase variance and therefore $t^{DP}$ is less strongly concentrated than $t^2$, leading to excessive transgressions of the threshold $q_{1-\alpha}$. Consequently, to guarantee an accurate approximation of the nominal level, a different threshold is necessary.\\
Hypothetically, if we knew the true distribution of $t^{DP}$ under $H_0$, we could analytically calculate the exact $\alpha$-quantile $q^{exact}_{1-\alpha}$ and use the rejection rule "$t^{DP}>q^{exact}_{1-\alpha}$". Of course, in practice, these  quantiles are not available, but we can use a \textit{parametric bootstrap} to approximate
$q^{exact}_{1-\alpha}$ by an empirical version $q^*_{1-\alpha}$ calculated from the data. In Algorithm \ref{B1} we describe the systematic derivation of $q^*_{1-\alpha}$. \\[1ex]

\begin{algorithm}
	\small
    \algorithmicrequire \; \parbox[t]{\dimexpr\linewidth-\algorithmicindent}{Covariance matrices: $\hat{\Sigma}_X^{DP}$, $\hat{\Sigma}_Y^{DP}$,\\ sample sizes: $n_1$,$n_2$, bootstrap iterations: $B$} \\[0.2cm]
    \algorithmicensure \, Empirical $1-\alpha$ quantile of $t^{DP}$: $ q_{1-\alpha}^*$.
 \begin{algorithmic}[1]
 \Function{\textnormal{QB}}{$\hat{\Sigma}_X^{DP}$, $\hat{\Sigma}_Y^{DP}$, $n_1$,$n_2$, $B$}
	\For {$i=1,\hdots,B$}
        \State Sample $\bar{X}^* \sim \mathcal{N}(0,\frac{\hat{\Sigma}_X^{DP}}{n_1})$ and $\bar{Y}^*\sim \mathcal{N}(0,\frac{\hat{\Sigma}_Y^{DP}}{n_2})$\For{$k=1,\hdots,d$}
	    \State Generate $Z_k \sim Lap(0, \frac{2 md}{n_1 (\ve/4)})$
        \State Generate $Z_k' \sim Lap(0, \frac{2 md}{n_2 (\ve/4)})$
        \EndFor
    \State Define $\bar{X}^{DP*} := \bar{X}^*+(Z_1,...,Z_d)$
    \State Define $\bar{Y}^{DP*} := \bar{Y}^*+(Z_1',...,Z_d')$
    \State Define
	\State  ${t^{DP}_i}^* := \frac{n_1n_2}{n_1+n_2} \norm{[\hat{\Sigma}^{DP}]^{-1/2}(\bar{X}^{DP*}-\bar{Y}^{DP*})}_2^2$
	\EndFor
	\State Sort statistics in ascending order: 
	\State ${(t^{DP}_{(1)}}^*,...,{t^{DP}_{(B)}}^*)= sort(({t^{DP}_{1}}^*,...,{t^{DP}_{B}}^*))$
	\State Define ${q}^*_{1-\alpha}:={{t^{DP}}^*_{(\lfloor(1-\alpha)B\rfloor)}}$
	\State \Return ${q}^*_{1-\alpha}$
	\EndFunction
    \end{algorithmic}
	\caption{Quantile Bootstrap (QB)}\label{B1}
\end{algorithm}

Algorithm \ref{B1} creates $B$ \textit{bootstrap versions}  ${t_1^{DP}}^*,...,{t_B^{DP}}^*$, that mimic the behavior of $t^{DP}$. So e.g. $\bar X^{DP*}$ (in ${t_i^{DP}}^*$) has a distribution close to that of $\bar X^{DP}$ (in $t^{DP}$), which, if centered, is approximately normal with covariance matrix $\Sigma_X/n_1$. As a consequence of this parallel construction, the \textit{empirical $1-\alpha$-quantile} $q^*_{1-\alpha}$ is close to the true $(1-\alpha)$-quantile of the distribution of $t^{DP}$, at least if the number  $B$ of bootstrap replications is sufficiently large. In practice, the choice of $B$ depends on $\alpha$ (where small $\alpha$ require larger $B$), 
but our simulations suggest that for a few hundred 
iterations the results are already reasonable even for nominal levels as small as $ 1\%$.

\begin{Thm}\label{Theorem_bootstrap}
The decision rule "reject if 
\begin{equation}
\label{hd2}
t^{DP}>q_{1-\alpha}^*
\end{equation}
(Algorithm \ref{alg_test})", where $q_{1-\alpha}^*$ chosen by Algorithm \ref{B1},
yields a consistent, asymptotic level-$\alpha$ test in the sense that
$$
    \lim_{B \to \infty} \lim_{n_1, n_2 \to \infty} \mathbb{P}_{H_0}(t^{DP}>q_{1-\alpha}^*) = \alpha,
$$
(level $\alpha$) and
$$
    \lim_{n_1, n_2 \to \infty} \mathbb{P}_{H_1}(t^{DP}>q_{1-\alpha}^*) = 1
$$
(consistency).
\end{Thm}

\section{Simulation} \label{Sec_4}

In this Section we investigate the empirical properties of our methodology by means of a small simulation study. \\
\textbf{Data generation:} In the following, the first sample $X_1,...,X_{n_1}$ is drawn from the uniform distribution on the $d$-dimensional cube $[-\sqrt{3}, \sqrt{3}]^d$, whereas the second sample $Y_1,...,Y_{n_2}$ is uniformly drawn from the shifted cube $[-\sqrt{3}+a/\sqrt{d}, \sqrt{3}+a/\sqrt{d}]^d$. Here $a\ge 0$ determines the mean difference of the two samples. In particular $a=0$ corresponds to the hypothesis $\mu_X = \mu_Y = (0,...,0)^T$, whereas for $a>0$, $\norm{\mu_X-\mu_Y}_2=a$. 
We also point out that both samples have the same covariance matrix $\Sigma_X = \Sigma_Y =  Id_{d \times d}$. As a consequence deviations in each component of $\mu_X-\mu_Y$ have equal influence on the rejection probability. 
\\In addition to the above model, we have also considered non-trivial covariance structures. The corresponding simulations are detailed in the Appendix (but the results are similar).
\\\textbf{Parameter settings:} In the following we discuss various settings: We consider different group sizes $n$, between $10^2$ and $10^5$, privacy levels $\ve=1/10,1/2,1,5$ and dimensions $d=1,10,30$. The nominal level $\alpha$ is fixed at $5\%$ and the number of bootstrap samples is consistently $B=200$. All below results are based on $1000$ simulation runs.\\
\textbf{Empirical type-1-error:} We begin by studying the behavior of our test decisions under the null hypothesis ($a=0$). In Table \ref{tab:1} we report the empirical rejection probabilities for the bootstrap test
\eqref{hd2} (top) and the asymptotic test  \eqref{hd1} (bottom). The empirical findings confirm  our theoretical results from the previous Section. 

\begin{table*}[!]
\caption {Empirical type-1-error} \label{tab:1} 
\resizebox{17cm}{!}{
  \begin{tabular}{c|c| c c c c | c c c c | c c c c }
    && \multicolumn{4}{ c |}{$d=1$}&\multicolumn{4}{ c |}{$d=10$}&\multicolumn{4}{c}{$d=30$}\\\hline
    
    &\backslashbox{$\ve$}{$n_1=n_2$}&$10^2$&$10^3$&$10^4$&$10^5$&$10^2$&$10^3$&$10^4$&$10^5$&$10^2$&$10^3$&$10^4$&$10^5$  \\ \hline \hline
   \multirow{3}{*}{\begin{sideways}test \eqref{hd2} \end{sideways}}
      &$0.1$&$0.052$&$0.046$&$0.051$&$0.048$&$0.058$&$0.05$&$0.068$&$0.063$&$0.04$&$0.057$&$0.056$&$0.062$\\
   &$0.5$&$0.054$&$0.05$&$0.059$&$0.05$&$0.039$&$0.06$&$0.057$&$0.052$&$0.054$&$0.054$&$0.06$&$0.056$\\
    &$1$&$0.053$&$0.05$&$0.054$&$0.053$&$0.048$&$0.061$&$0.038$&$0.069$&$0.048$&$0.063$&$0.056$&$0.054$\\
    &$5$&$0.041$&$0.053$&$0.043$&$0.053$&$0.055$&$0.053$&$0.056$&$0.051$&$0.044$&$0.05$&$0.062$&$0.052$\\
    \hline\hline
      \multirow{3}{*}{\begin{sideways}test \eqref{hd1} \end{sideways}}
     &$0.1$&$0.738$&$0.676$&$0.328$&$0.093$&$1$&$1$&$1$&$1$&$1$&$1$&$1$&$1$\\
      &$0.5$&$0.4$&$0.154$&$0.055$&$0.058$&$1$&$1$&$1$&$0.891$&$1$&$1$&$1$&$1$\\
    &$1$&$0.24$&$0.063$&$0.057$&$0.044$&$1$&$1$&$0.993$&$0.428$&$1$&$1$&$1$&$1$\\
    &$5$&$0.054$&$0.047$&$0.045$&$0.039$&$0.990$&$0.933$&$0.181$&$0.062$&$1$&$1$&$0.999$&$0.301$\\
  \end{tabular}}
  
  \caption*{\\Empirical type-1-error of the asymptotic test \eqref{hd1} (bottom) and the bootstrap test \eqref{hd2}
  (top) for various privacy parameter $\ve=0.1,0.5,1,5$, sample sizes $n_1=n_2=10^2,10^3,10^4,10^5$ and dimensions $d=1,10,30$.}
\end{table*}
\smallskip
\begin{figure*}[!]
\centering
\begin{subfigure}[c]{.45\linewidth}
\caption{\textbf{$\ve=0.1$}}
\includegraphics[width=1\linewidth]{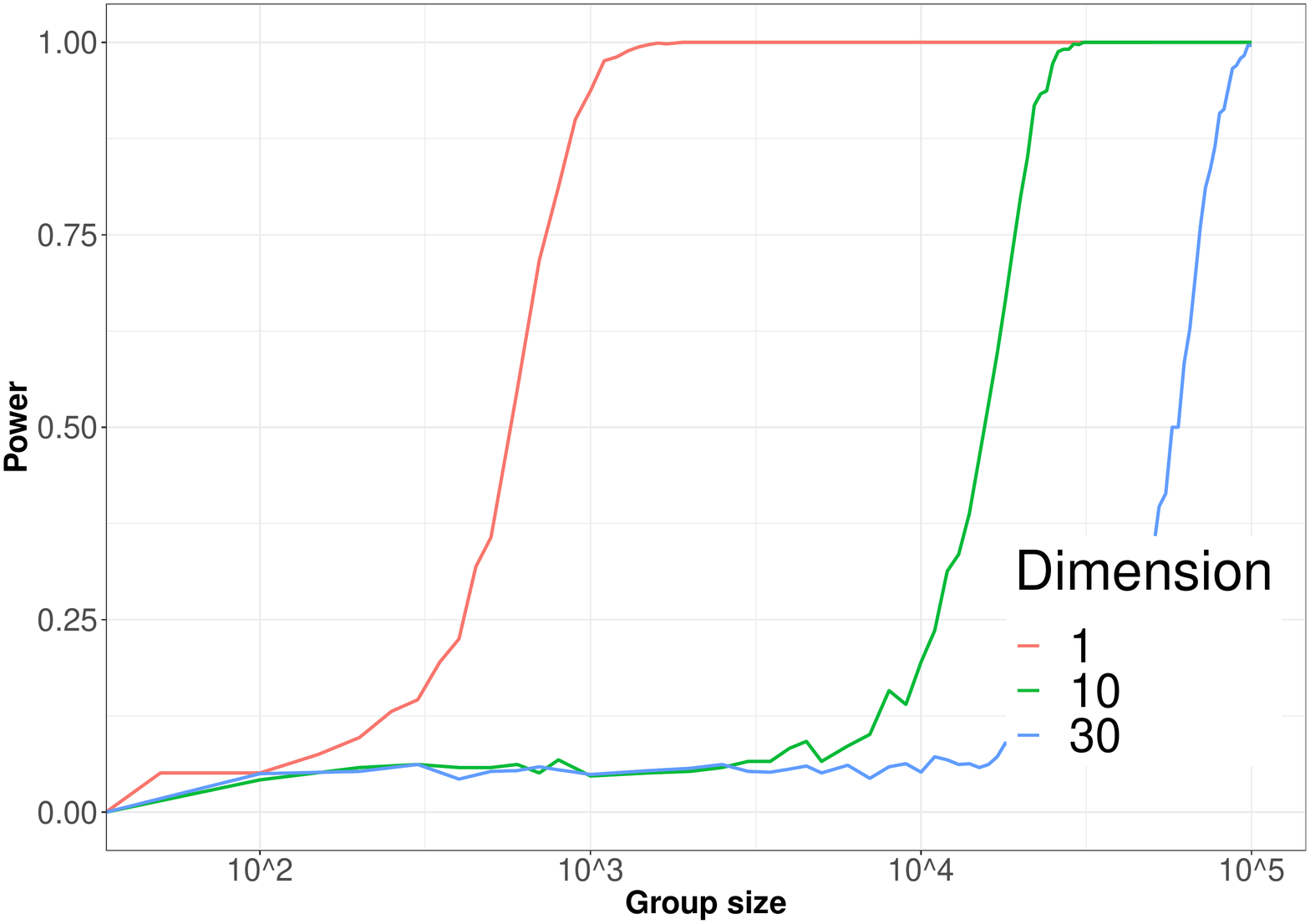}
\label{fig:eps0.1}
\end{subfigure}%
\quad
\begin{subfigure}[c]{.45\linewidth}
\caption{\textbf{$\ve=0.5$}}
\includegraphics[width=1\linewidth]{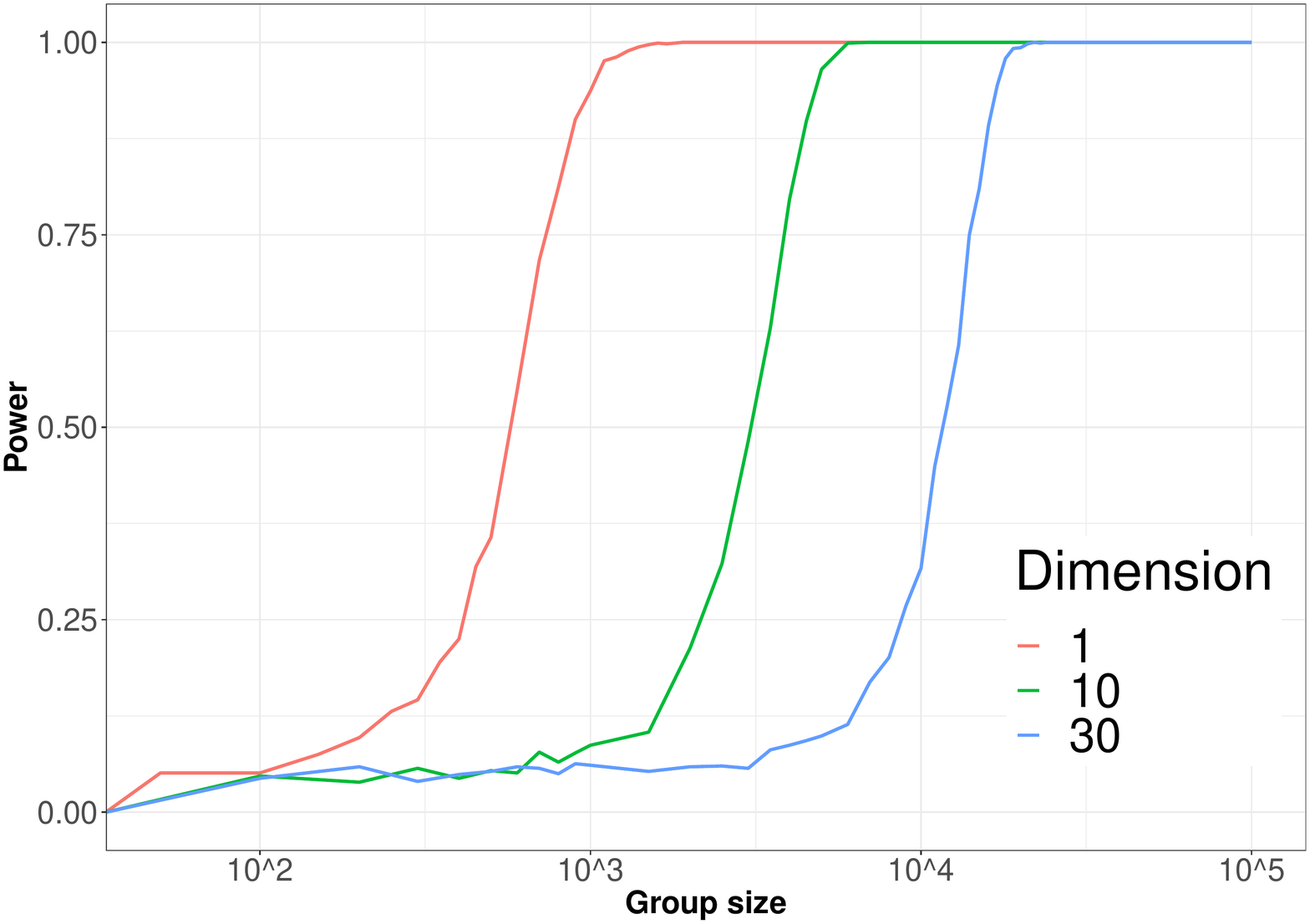}
\label{fig:eps0.5}
\end{subfigure}%

\begin{subfigure}[c]{.45\linewidth}
\caption{\textbf{$\ve=1$}}
\includegraphics[width=1\linewidth]{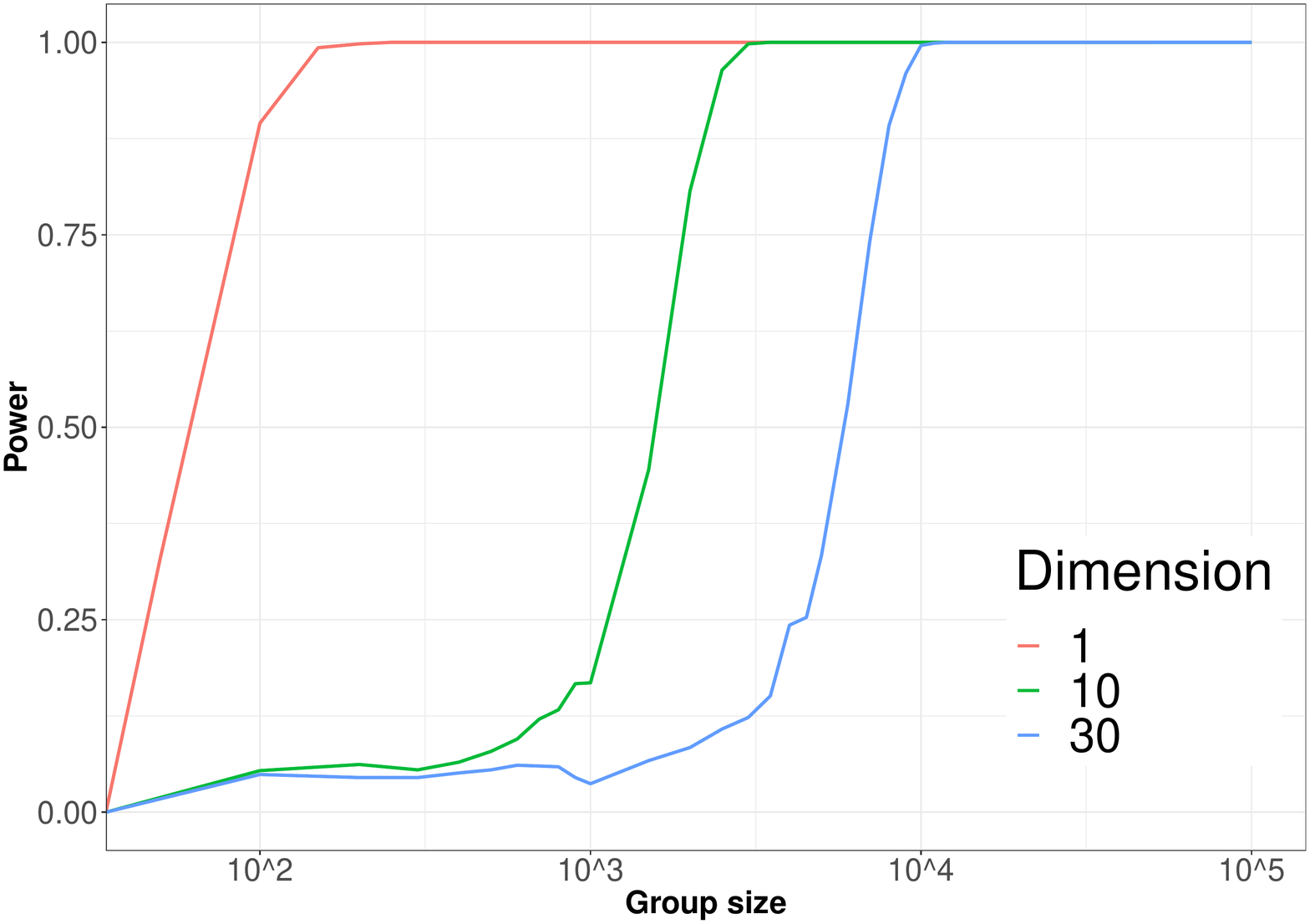}
\label{fig:eps1}
\end{subfigure}%
\quad
\begin{subfigure}[c]{.45\linewidth}
\caption{\textbf{$\ve=5$}}
\includegraphics[width=1\linewidth]{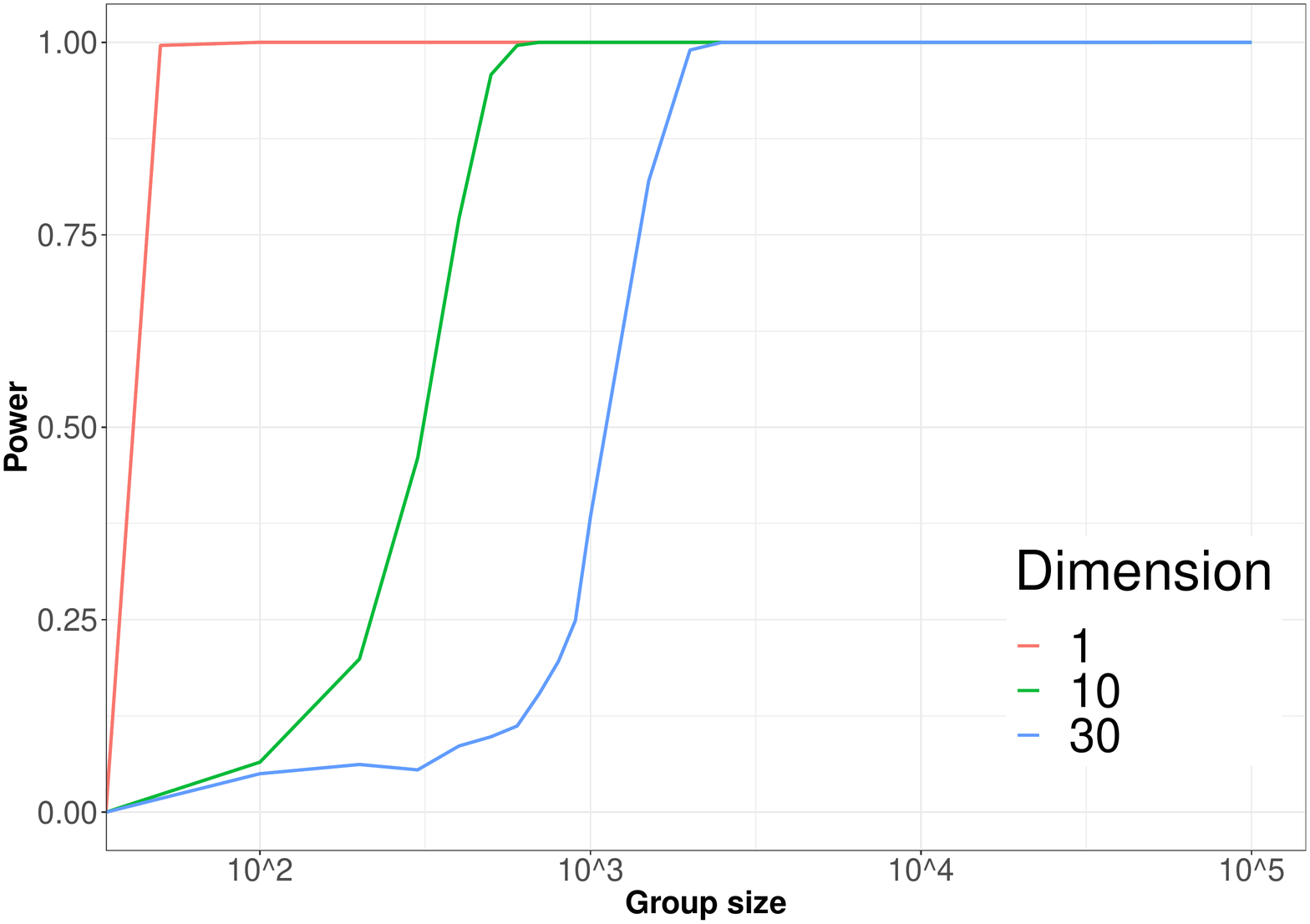}
\label{fig:eps5}
\end{subfigure}%
 \caption{Simulated power of the bootstrap test \eqref{hd2} under a   uniform alternative for $\ve=0.1,0.5,1,5$  and different group sizes.\label{Fig_power}}
\end{figure*}

On the one hand we  observe that the bootstrap test approximates the nominal-level reasonably well (compare Theorem \ref{Theorem_bootstrap}), even in scenarios with small sample size and high dimensions. In contrast the validity of the asymptotic test  \eqref{hd1} 
depends on the negligibility of privatization effects (see discussion of Theorem \ref{Theorem_asymptotic}). 
Consequently it works best for large $\ve$ and large sample  sizes. However, for higher dimensions $d$, the asymptotic approach breaks down quickly, in the face of more noise by privatizations and thus stronger digressions from the limiting distribution.\\
\textbf{Empirical power:} Next we consider the power of our test. Given the poor performance of the asymptotic test \eqref{hd1}  in higher dimensions (the key interest of this paper) we restrict our analysis on the bootstrap test
\eqref{hd2} for the sake of brevity. In the following we consider the alternative for $a=1$. Recall that $\norm{\mu_X-\mu_Y}_2=a$ is independent of the dimension. However, we expect more power in low dimensions due to weaker privatization. In Figure \ref{Fig_power}, we display a panel of empirical power curves, each graphic reflecting a different choice of the privacy parameter ($\varepsilon=1/10,1/2,1,5$) and each curve corresponding to a different dimension ($d=1,10,30$). The group size is reported in logarithmic scale on the $x$-axis and the rejection probability on the $y$-axis. As might be expected, low dimensions and weak privatizations (i.e. large $\ve$) are directly associated with a sharper increase of the power curves and smaller sample sizes to attain high power. For instance, moving from $\varepsilon=1/2$ (high privatization) to the less demanding $\varepsilon=5$ (low privatization) means that a power of $90\%$ is attained with group sizes that are about an order of magnitude smaller. Similarly, increasing dimension translates into lower power: To attain for $\varepsilon=0.1$ and $d=30$, high power requires samples of a few ten thousand observations (see Figure \ref{Fig_power}(\subref{fig:eps0.1})). Even though such numbers are not in excess of those used in related studies (see e.g. \cite{ding2018comparing}) nor of those raised by large tech cooperations, this trend indicates that comparing means of even higher dimensional populations might require (private) pre-processing to reduce dimensions. 

\section{Conclusion} \label{Sec_5}

In this paper, we have considered a new way to test multidimensional mean differences under the constraint of differential privacy. Our test employs a privatized version of the popular Hotelling's $t^2$-statistic, together with a bootstraped rejection rule. While strong privacy requirements always go hand in hand with a loss in power, the test presented in this paper respects the nominal level $\alpha$ with high precision, even for moderate sample sizes, high dimensions and strong privatizations. The empirical advantages are underpinned by theoretical guarantees for large samples. Given the easy implementation and reliable performance the test can be used as an automatized part of larger analytical structures.

\section*{Acknowledgement}
This work was partially funded by the DFG under Germany's Excellence Strategy - EXC 2092 CASA - 390781972.

\newpage $ $ \newpage

\newpage

\phantomsection
\addcontentsline{toc}{section}{Bibliography}

\setlength{\bibsep}{1.3pt}

\begin{small}

\bibliography{literature.bib}

\end{small}

\onecolumn

\appendix
\section{Proofs and technical details}
The Supplement is dedicated to the proofs of the theoretical results presented in Section 3. In the following we use the notation "$\overset{d}{\to}$" to denote weak convergence and "$\overset{\PR}{\to}$" to denote convergence in probability. Moreover consider a generic sequence of real valued random variables $(X_n)_{n \in \mathbb{N}}$ and a deterministic sequence of positive real numbers $(a_n)_{n \in \mathbb{N}}$. $(X_n)_{n \in \mathbb{N}}$ is of order $\mathcal{O}_{\PR}(a_n)$ (we write "$X_n = \mathcal{O}_{\PR}(a_n)$"), if for any $\delta>0$ there exist $M=M(\delta)>0$ and $n_0=n_0(\delta, M) \in \N$, s.t. $\PR(| X_n|/a_n>M)<\delta$ for all $n\geq n_0$. Similarly $(X_n)_{n \in \mathbb{N}}$ satisfies $o_{\PR}(a_n)$ (we write "$X_n = o_{\PR}(a_n)$"), if for any (arbitrarily small) $\delta>0$, we have $\lim_{n\to\infty}\PR(|X_n|/a_n\geq\delta)=0$.

For further details of these terms see \cite{bishop}.

\subsection{Proof of Theorem 3.3}

Applying the composition theorem of DP (see \cite{10.1561/0400000042}, p.43) on all four $\ve/4$-\textit{differentially private} algorithms (two times Laplace-noise and two times ED-algorithm) yields that the vector $V^{DP}:=(\bar{X}^{DP}, \bar{Y}^{DP},\hat\Sigma_X^{DP}, \hat\Sigma_Y^{DP})$ is $\ve$-\textit{differentially private}. As $t^{DP}$ is a measurable transformation of $V^{DP}$, $t^{DP}$ remains $\varepsilon$-DP according to the stability of DP against post-processing (again see \cite{10.1561/0400000042}, p.19).

\subsection{Proof of Theorem 3.4}
We begin by investigating $t^{DP}$ under the null hypothesis. 
Recall that by definition $\bar{X}^{DP}=\bar{X}+\mathcal{O}_\PR(1/n), \bar{Y}^{DP}=\bar{Y}+\mathcal{O}_\PR(1/n)$. This together with
 the multivariate central limit theorem (see Example 2.18 \cite{van2000asymptotic}, p.16) implies that under the null hypothesis ($H_0: \mu_X = \mu_y$), we have
	\begin{align} \label{eq_normality}
		\sqrt{\frac{n_1n_2}{n_1+n_2}} (\bar{X}^{DP}-\bar{Y}^{DP})
		\overset{d}{\longrightarrow} M \sim \mathcal{N}_d\left(0, \xi_2\Sigma_X+\xi_1\Sigma_Y\right) \overset{d}{=} \mathcal{N}_d\left(0, \Sigma_X\right)~.
	\end{align}
Here ${\cal N}_d ( \nu , \Sigma )$ denotes a $d$-dimensional normal distribution with mean vector
	$\nu$ and covariance matrix $\Sigma$ and
	"$\overset{d}{=}$" denotes equality in distribution, which follows because $\xi_2\Sigma_X+\xi_1\Sigma_Y=\Sigma_X$ according to Assumptions (2) and (3).
	Next, using that $\hat\Sigma_X^{DP}$ and $\hat\Sigma_Y^{DP}$ are consistent estimators for $\Sigma_X=\Sigma_Y$ (see  Section \ref{Subsec_con}), we can conclude that
	\begin{equation*}
	\hat{\Sigma}^{DP}=\frac{(n_1-1)\hat{\Sigma}_{X}^{DP}+(n_2-1)\hat{\Sigma}_{Y}^{DP}}{n_1+n_2-2}
	\end{equation*}
	is also a consistent estimator for $\Sigma_X$ since
	\begin{equation*}
		\hat{\Sigma}^{DP} \overset{\PR}{\longrightarrow} \xi_1 \Sigma_X+\xi_2 \Sigma_Y= \Sigma_X~.
	\end{equation*}
	By the continuous mapping theorem (see Theorem 2.3 \cite{van2000asymptotic}, p.7), we have
	\begin{equation}\label{eq_consistency}
			\hat\Sigma^{-1}\overset{\PR}{\longrightarrow} \Sigma_X^{-1}~.
	\end{equation}
	
	Here we have used that $\Sigma_X$ is invertible according to Assumption (2). Combining \eqref{eq_normality} and \eqref{eq_consistency} yields 

	\begin{align*}
		t^{DP}&= \frac{n_1n_2}{n_1+n_2} (\bar{X}^{DP}-\bar{Y}^{DP})^T[{\hat{\Sigma}^{DP}}]^{-1}(\bar{X}^{DP}-\bar{Y}^{DP})
		\\&=\frac{n_1n_2}{n_1+n_2} (\bar{X}^{DP}-\bar{Y}^{DP})^T\left[(\hat{{\Sigma}}^{DP})^{-1}-\Sigma_X^{-1}\right](\bar{X}^{DP}-\bar{Y}^{DP})
		\\&+\frac{n_1n_2}{n_1+n_2} (\bar{X}^{DP}-\bar{Y}^{DP})^T\left[\Sigma_X^{-1}\right](\bar{X}^{DP}-\bar{Y}^{DP})
		\\&\overset{d}{\longrightarrow}0+ M^T\Sigma_X^{-1}M  \overset{d}{=}  \chi_d^2~,
	\end{align*}

	where again $Z\sim \mathcal{N}_d\left(0,\Sigma_X\right)$ and $\chi_d^2$ denotes a chi-square distribution with $d$ degrees of freedom. The last equality follows from e.g. \cite{van2000asymptotic} p.242. 

	This convergence already entails asymptotic level $\alpha$ of the rejection rule "reject if $t^{DP}>q_{1-\alpha}$".\\
	Next assume that $H_1$ holds with $\mu_X-\mu_Y=a \neq 0$. Then we have to prove that
\begin{equation} \label{eq_consistency_test}
    \lim_{n_1,n_2\to\infty} \PR_{H_1}(t^{DP}\ge q_{1-\alpha})=1.
\end{equation}
According to the first step $\bar{X}^{DP}=\bar{X}+\mathcal{O}_\PR(1/n), \bar{Y}^{DP}=\bar{Y}+\mathcal{O}_\PR(1/n)$, which together with the law of large numbers implies $\bar{X}^{DP} = \mu_X+o_\PR(1)$, $\bar{Y}^{DP} = \mu_Y+o_\PR(1)$.  As a consequence
\begin{align*}
    t^{DP} = \frac{n_1n_2}{n_1+n_2}\norm{{\Sigma^{DP}}^{-1/2}(\bar{X}^{DP}-\bar{Y}^{DP})}_2^2 = \frac{n_1n_2}{n_1+n_2}\norm{\Sigma^{-1/2}a }_2^2 +o_\PR \Big( \frac{n_1n_2}{n_1+n_2}\Big)~.
\end{align*}
The deterministic term on the right dominates (since $\|\Sigma^{-1/2}a \|_2^2>0$) and consequently the right diverges to $\infty$ in probability. This directly implies \eqref{eq_consistency_test}.

\subsection{Proof of Theorem 3.5}
We begin by noticing two facts:
\begin{itemize}
    \item[1)] Recalling the definition
    of $c_1$ and $c_2$ in the main part of the paper, it follows that
      $diag(c_1), diag(c_2) \to  \mathbf{0}_{d \times d}$ ($\mathbf{0}_{d \times d}$ denotes the $d\times d$ matrix with only 0 as entries), $\hat\Sigma_X^{DP}, \hat\Sigma_Y^{DP}, \hat\Sigma^{DP} \overset{\PR}{\to}\Sigma_X$ and $\frac{n_1}{n}\to \xi_1$,$\frac{n_2}{n}\to\xi_2$  as $n_1, n_2 \to \infty$.
    \item[2)] Denoting the bootstrap statistic in the $i$-th iteration ${t_{i}^{DP}}^*$, we can express it as 
    \begin{align*}
       {t_{i}^{DP}}^*\overset{d}{=} \frac{n_1n_2}{n_1+n_2} &\Big\|\left(diag(c_1+c_2)+\hat\Sigma^{DP}\right)^{-1/2}\\
       & \times\left(\tilde X^{(i)} [\hat\Sigma_X^{DP}/n_1]^{1/2}+Z^{(i)}-\tilde Y^{(i)} [\hat\Sigma_Y^{DP}/n_2]^{1/2}+Z^{'(i)}\right)
         \Big\|_2^2 =: \norm{G(i)}_2^2
    \end{align*}
    where $\tilde X^{(i)}, \tilde Y^{(i)}$ are $d$-dimensional, standard normal random variables, independent of each other and everything else and $G(i)$ is defined in the obvious way. Notice that the (super-)index $(i)$ expresses the dependence on $i=1,...,B$.
\end{itemize}
 We now focus on the weak convergence of $G(i)$ for a fixed index $i$ and growing sample sizes. Given that $\Sigma_X=\Sigma_Y$ is invertible, it follows by fact 1) and simple calculations that $G(i) = G'(i)+o_\PR(1)$, where 
\begin{equation}
    G'(i) :=  (\xi_2^{1/2}\tilde X^{(i)} \Sigma_X^{1/2}-\xi_1^{1/2}\tilde Y^{(i)} \Sigma_X^{1/2})\Sigma_X^{-1/2} = \xi_2^{1/2}\tilde X^{(i)} -\xi_1^{1/2}\tilde Y^{(i)}. 
\end{equation}

By construction $G'(i)$ follows a standard normal distribution and is independent of all $G'(j)$ for $i \neq j$ as well as $t^{DP}$. As a consequence we have weak convergence $(t^{DP}, \norm{G(1)}_2^2, ..., \norm{G(B)}_2^2) \overset{d}{\to} (T, T_1,...,T_B)$, where $T, T_1,...,T_B$ are independent identically distributed copies of a $\chi_d^2$ -distribution. Now let $q_{1-\alpha}^*$ be the empirical $(1-\alpha)$-quantile, based on $T_1,...,T_B$. It holds that
$$
    \lim_{n_1, n_2 \to \infty} \mathbb{P}_{H_0}({t^{DP}}>q_{1-\alpha}^*) = \mathbb{P}(T >q_{1-\alpha}^{*}).
$$
It follows that as $B \to \infty$ it holds that $q_{1-\alpha}^{*} \overset{\PR}{\to} q_{1-\alpha}$ (the true upper $\alpha$-quantile of the distribution of $T$), e.g. according to Example 3.9.21 in \cite{van1996weak} . This completes the proof.

\subsection{Consistency of the private covariance}\label{Subsec_con}

\begin{Prop}
Under Assumptions (1)-(3) it holds for any $\varepsilon>0$ that 
$$
\hat \Sigma_X^{DP} \overset{\PR}{\to} \Sigma_X, \quad \textnormal{and} \quad \hat \Sigma_Y^{DP} \overset{\PR}{\to} \Sigma_Y.
$$
\end{Prop}
     We only conduct the proof for $\Sigma_X$ (the proof for the second sample follows analogously). Let $\|\cdot \|_F$ denote the Frobenius-norm. According to Theorem 2 in \cite{NEURIPS2019_4158f6d1}, it follows with $\beta>0$ and $b_{n_1} := \frac{\sqrt{d}(d+1)\log2dn_1}{\ve}$ that
\begin{equation*}
	\PR\left(\norm{\hat\Sigma_X-\hat\Sigma^{DP}_X}_F\leq \frac{1}{n_1}K_1\left(\sqrt{\sum_{i=1}^{d} \frac{\lambda_i(n_1\hat\Sigma_X)(d+1)}{\ve}\left(d\log\lambda_1(n_1\hat\Sigma_X)+\log\frac{1}{\beta/(2d)}\right)}+b_{n_1}\right)\middle|X\right)\geq 1-\beta~.
\end{equation*}
Here we have defined $X:=(X_1,...,X_{n_1})$, to abbreviate conditioning on the whole sample and $\lambda_1(n_1\hat\Sigma_X)\geq\hdots\geq \lambda_d(n_1\hat\Sigma_X)$ denote the ordered eigenvalues. 
Now choosing $\beta=\frac{1}{n_1}$, we obtain
\begin{equation*}
	\PR\left(\norm{\hat \Sigma_X-\hat\Sigma_X^{DP}}_F\leq \frac{1}{n_1}K_1\left(\sqrt{\sum_{i=1}^{d} \frac{\lambda_i(n_1\hat\Sigma_X)(d+1)\left(d\log\lambda_1(n_1\hat\Sigma_X)+\log2d n_1\right)}{\ve}}+b_{n_1}\right)\middle|X\right)\geq 1-\frac{1}{n_1}~.
\end{equation*} 
Thus, we end up with
\begin{equation*}
	\PR\left(\norm{\hs_X-\hat\Sigma_X^{DP}}_F\leq \frac{1}{\sqrt{n_1}}K_1\left(\sqrt{\sum_{i=1}^{d} \frac{\lambda_i(n_1\hat\Sigma_X)(d+1)\left(d\log\lambda_1(n_1\hat\Sigma_X)+\log2d n_1\right)}{n_1\ve}}+\frac{b_{n_1}}{\sqrt{n_1}}\right)\middle|X\right)\geq 1-\frac{1}{n_1}~.
\end{equation*}
Since $\frac{\lambda_i(n_1\hat\Sigma_X)}{n_1}=O_{\PR}(1)$, $\frac{b_{n_1}}{\sqrt{n_1}}\overset{n_1\to\infty}{\to}0$, we obtain the consistency if we note that 
\begin{align*}
    &\lim_{n\to\infty}	\PR\left(\norm{\hs_X-\hat\Sigma_X^{DP}}_F\leq \frac{1}{{n_1}^{1/4}}K_1\left(\sqrt{\sum_{i=1}^{d} \frac{\lambda_i(n_1\hat\Sigma_X)}{n_1}\frac{(d+1)\left(d\log\lambda_1(n_1\hat\Sigma_X)+\log2d n_1\right)}{\sqrt{n_1}\ve}}+\frac{b_{n_1}}{\sqrt{n_1}}\right)\middle|X\right)\\ &\geq \lim_{n\to\infty} 1-\frac{1}{n_1}=1~.
\end{align*}
Now defining $$
R_n:=\frac{1}{{n_1}^{1/4}}K_1\left(\sqrt{\sum_{i=1}^{d} \frac{\lambda_i(n_1\hat\Sigma_X)}{n_1}\frac{(d+1)\left(d\log\lambda_1(n_1\hat\Sigma_X)+\log2d n_1\right)}{\sqrt{n_1}\ve}}+\frac{b_{n_1}}{\sqrt{n_1}}\right)
$$ 
and noting that $R_n \overset{\PR}{\to}0$  we obtain that
\begin{align*}
    \lim_{n\to\infty}\PR\left(\norm{\hs_X-\hs_X^{DP}}_F\leq R_n\right)=\lim_{n\to\infty}\int  \PR\left(\norm{\hs_X-\hs_X^{DP}}_F\leq R_n\middle|X\right) d\PR^{X}\geq
    \lim_{n\to\infty}\int  (1-\frac{1}{n}) d\PR^{X}=
    1.
\end{align*}
Recalling that $\hat \Sigma_X$ is consistent for $\Sigma_X$ now yields the desired result.
\newpage
\section{Algorithms}
In the following we will state two algorithms which describe the covariance privatization. Here, Algorithm \ref{algpriv} \textbf{ED} is used for the privatization while Algorithm \ref{alg_expo} describes the eigenvector sampling process. 
\\In Algorithm \ref{algpriv} \textbf{ED} the privatization budget is not supposed to be separated (for eigenvalues and eigenvectors) in the case $d=1$ (as eigenvector privatization is unnessecary for $d=1$). For convenience we leave the Algorithm in that shape. Also note that we have bounded the $\ell_2$ norm of the data vector in Algorithm \ref{algpriv}, which results in more convenient sensitivity calculation for both the Laplace- and Exponential Mechanism (for details see \cite{NEURIPS2019_4158f6d1})
\begin{algorithm}
\small
    \algorithmicrequire \; \parbox[t]{\dimexpr\linewidth-\algorithmicindent}{$\hat C\in \R^{d\times d}$, privacy parameter $\ve$, sample size $n$} \\[0.2cm]
    \algorithmicensure \, Privatized covariance matrix $\hat\Sigma^{DP}$
	\begin{algorithmic}[1] 
	\State Separate the privacy budget uniformly in $d+1$ parts, i.e. each step $\frac{\ve}{d+1}$
    \Function{ED}{$\hat C$,$\ve$,$n$}
		\State Initialize $C_1:=\frac{n\hat C}{d m^2}$, $P_1:=I_d$. \State Privatize the eigenvalue vector by $(\bar{\lambda}_1,\hdots,\bar{\lambda}_d)^T=\left|(\hat{\lambda}_1,\hdots,\hat{\lambda}_d)^T+\left(Lap\left(\frac{2}{(\ve/(d+1))}\right),\hdots,Lap\left(\frac{2}{(\ve/(d+1)) }\right)\right)^T\right|$. 
		\For{$i=1,\hdots,d-1$}
		\State Sample $\bar{u}_i \in S^{d-i}$ with $\bar{u}_i:=Sample(\hat C, \frac{\ve}{d+1})$ and let $\bar{v}_i :=P_i^T\bar{u}_i$.
		\State Find an orthonormal basis $P_{i+1}\in\R^{(d-i)\times d}$ orthogonal to $\bar{v}_1, \hdots, \bar{v}_i$. 
		\State Let $C_{i+1}:=P_{i+1}\hat C P_{i+1}^T\in \R^{(d-i)\times(d-i)}$.
		\EndFor
		\State Sample $\bar{u}_d \in S^0$ proportional to $f_{C_d}(u)=\exp\left((\frac{\ve_i}{4}) u^TC_du \right)$ and let $\bar{v}_d :=P_d^T\bar{u}_d$.
		\State $C^{ED}:=\sum_{i=1}^{d} \bar{\lambda}_i\bar{v}_i\bar{v}_i^T$.
		\State \Return $\hat\Sigma^{DP}=\frac{1}{n}C^{ED}$
	 \EndFunction 
    \end{algorithmic}
	\caption{Covariance estimation with algorithm \textbf{ED}}\label{algpriv}
\end{algorithm}

\begin{algorithm}
\small
    \algorithmicrequire \; \parbox[t]{\dimexpr\linewidth-\algorithmicindent}{$\tilde C\in \R^{q\times q}$, privacy parameter $\ve$} \\[0.2cm]
    \algorithmicensure \, Eigenvector $u$.
\begin{algorithmic}[1]
\Function{Sample}{$\tilde C$,$\ve$}
\State Define $A:= -\frac{\ve}{4}\tilde C+\frac{\ve}{4}\hl_1I_q$, where $\hl_1$ denotes the largest eigenvalue of $C$.
		\State Define $\Omega=I_q+2A/b$, where $b$ satisfies $\sum_{i=1}^{q}\frac{1}{b+2\lambda_i(A)}=1$.
		\State Define $M:=\exp(-(q-b)/2)(q/b)^{q/2}$~.
		\State Set $ANS=0$
		\While {$ANS =0$}
			\State Sample $X\sim \mathcal{N}_q\left(0, \Omega^{-1}\right)$ and set $u:=z/\norm{z}_2$. 
			\State With probability $\frac{\exp(-u^TAu)}{M(u^T\Omega u)^{q/2}}$ $ANS=1$ \State \Return $u$.
		\EndWhile
	\EndFunction
\end{algorithmic}
	\caption{Eigenvector sampling}\label{alg_expo}
\end{algorithm}
\newpage
\section{Additional simulations}
In the following we consider data with non-diagonal covariance matrix. Specifically, we consider the tridiagonal Toeplitz matrix given by
\begin{equation*}
    \Sigma_{T}= \begin{pmatrix}l&b&&&0\\b&l&b\\&\ddots&\ddots&\ddots\\&&b&l&b\\0&&&b&l\end{pmatrix} \in \R^{d\times d}~,
\end{equation*}
with diagonal entries $l=1$ and off diagonal entries $b=1/3$. Our data samples are now created as follows: Let $\tilde X_1,...,\tilde X_{n_1}$ be independently, uniformly drawn  from $[-\sqrt{3}, \sqrt{3}]^d$, whereas the $\tilde Y_1,...,\tilde Y_{n_2}$ are independently, uniformly drawn from the shifted cube $[-\sqrt{3}+a/\sqrt{d}, \sqrt{3}+a/\sqrt{d}]^d$. Now the data samples consist of the observations $X_i := \Sigma_T \tilde X_i$ for $i=1,...,n_1$ and $Y_j = \Sigma_T \tilde Y_j$ for $j=1,...,n_2$. Notice that the first sample is centered ($\mu_X = (0,...,0)^T$), whereas the second sample has expectation $\mu_Y = \Sigma_T (a,...,a)^T$. Both samples share the same covariance matrix $\Sigma_X = \Sigma_Y = \Sigma_{T}^2$, which can be easily shown to be invertible. 
As the samples $\tilde X_1,...,\tilde X_{n_1}$ and $\tilde Y_1,...,\tilde Y_{n_2}$ are linearly transformed by $\Sigma_T$, so is their support and we have to recalculate $m$, s.t. the cube $[-m, m]^d$ includes it. A simple calculation shows that we can choose $m=(\sqrt{3}+a/\sqrt{d})(2b+l)$. Notice that $m$ is larger than in our previous simulations (see Section 4), which translates into more noise added to the data. In the subsequent simulations we choose the parameters $d=10,30$, $\ve=0.1,0.5,1$ and $a=0$ (hypothesis) and $a=1$ (alternative). We omit comparisons with the asymptotic test (Theorem 3.4), since the results are similar as before and also do not consider $d=1$ (because in one dimension all matrices are diagonal).
\begin{table*}[h]
\caption {Empirical type-1-error} 
\centering

\resizebox{13cm}{!}{
  \begin{tabular}{c| c c c c | c c c c }
    &\multicolumn{4}{ c |}{$d=10$}&\multicolumn{4}{c}{$d=30$}\\\hline
    \backslashbox{$\ve$}{$n_1=n_2$}&$10^2$&$10^3$&$10^4$&$10^5$&$10^2$&$10^3$&$10^4$&$10^5$  \\ \hline \hline
  $0.1$&$0.049$&$0.041$&$0.058$&$0.05$&$0.05$&$0.064$&$0.046$&$0.062$\\
  $0.5$&$0.059$&$0.055$&$0.051$&$0.049$&$0.067$&$0.06$&$0.048$&$0.051$\\
  $1$&$0.07$&$0.046$&$0.045$&$0.059$&$0.052$&$0.042$&$0.039$&$0.056$\\
  \end{tabular}}
 \caption*{\\Empirical type-1-error for the bootstrap test (see Theorem 3.5) for various privacy parameter $\ve=0.1,0.5,1$, sample sizes $n_1=n_2=10^2,10^3,10^4,10^5$ and dimensions $d=10,30$. \label{label_table}}
\end{table*}
\begin{figure*}[h]
\centering
\caption{$ $\label{label_power}}
\begin{subfigure}[c]{.45\linewidth}
\caption{\textbf{$d=10$}}
\includegraphics[width=1\linewidth]{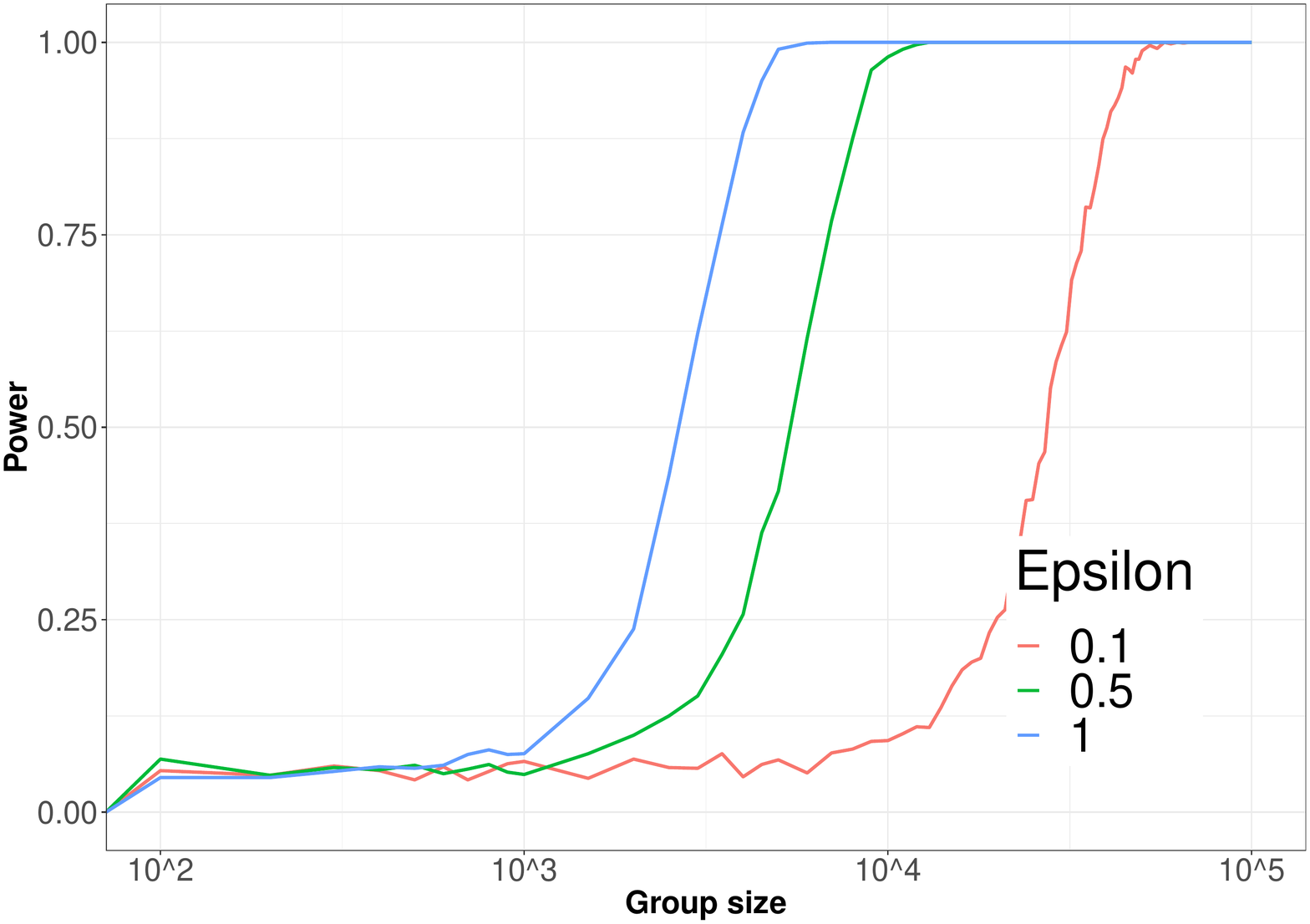}
\label{fig:eps0.1NEW}
\end{subfigure}%
\quad
\begin{subfigure}[c]{.45\linewidth}
\caption{\textbf{$d=30$} }
\includegraphics[width=1\linewidth]{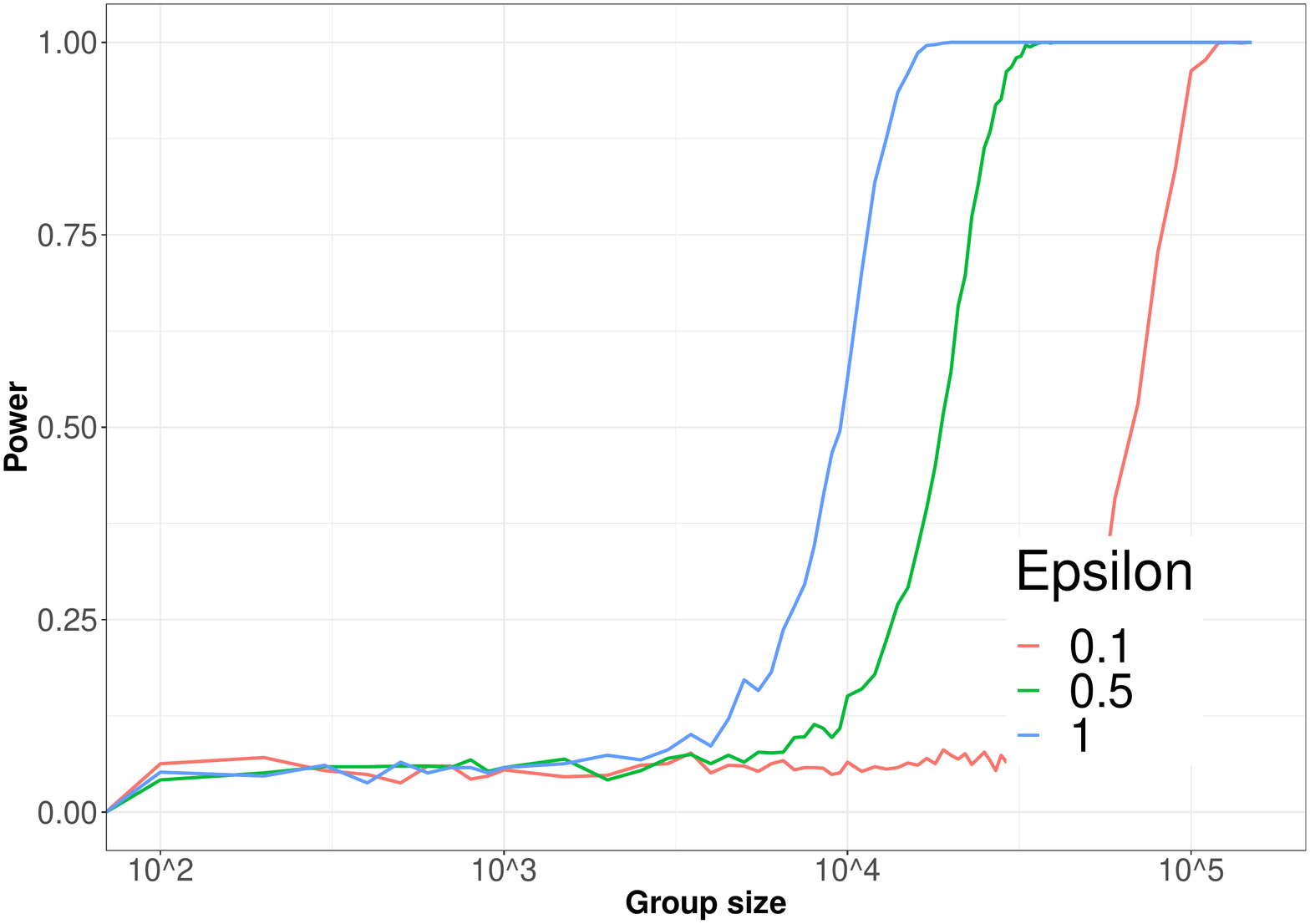}
\label{fig:eps0.5NEW}
\end{subfigure}%
\end{figure*}
Our results are reported in Table \ref{label_table} (under $H_0$) and in Figure \ref{label_power} (under $H_1$). The results under the hypothesis are practically identical to those in Section 4, where we have already observed good approximations of the nominal level. 
For the power curves we see slight differences, such as a slower increase, pointing at less power under the alternative. However this effect should not be attributed to the more complex covariance structure of the data, but rather to the stronger privatizations (larger $m$), which increase variance and hence reduce power.

\end{document}